\documentstyle[12pt]{article}

\makeatletter
\@addtoreset{equation}{section}

\newcommand{\nn}{\nonumber}
\newcommand{\vs}[1]{\vspace*{#1}}
\newcommand{\hs}[1]{\hspace*{#1}}

\newcommand{\p}{\partial}
\newcommand{\Half}{\frac12}
\newcommand{\unit}{\hbox to 3.8pt{\hskip1.3pt \vrule height 7.4pt
    width .4pt \hskip.7pt \vrule height 7.85pt width .4pt \kern-2.4pt 
    \hrulefill \kern-3pt \raise 3.7pt\hbox{\char'40}}}

\def\href#1#2{#2}

\begin{document}


\begin{titlepage}

\title{
\hfill\parbox{4cm}{
{\normalsize NSF-ITP-01-13}\\[-5mm]
{\normalsize SU-ITP-01/06}\\[-5mm]
{\normalsize\tt hep-th/0102173}
}
\\[40pt]
Branes Ending On Branes In A Tachyon Model
}
\author{
Koji {\sc Hashimoto}\thanks{{\tt 
koji@itp.ucsb.edu}}$\hspace{2mm}{}^a$
and 
Shinji {\sc Hirano}\thanks{{\tt 
hirano@itp.stanford.edu}}$\hspace{2mm}{}^b$
\\[10pt]
${}^a${\it Institute for Theoretical Physics,}\\
{\it University of California, Santa Barbara, CA 93106}\\
${}^b${\it Department of Physics, Stanford University,}\\
{\it Stanford, CA 94305}\\
}
\date{\normalsize February, 2001}
\maketitle
\thispagestyle{empty}

\begin{abstract}
\normalsize\noindent 
In a tachyon model proposed by Minahan and Zwiebach 
and derived in the boundary string field theory,
we construct various new solutions which correspond to nontrivial
brane configurations in string theory. Our solutions include
D$p$-D$(p\!-\!2)$ bound states, (F, D$p$) bound states,
string junctions, D$(p\!-\!2)$-branes ending on a D$p$-brane,
 D$(p\!-\!2)$-branes suspended between parallel D$p$-branes and their 
 non-commutative generalizations.
We find the Bogomol'nyi bounds and the BPS equations for some of our 
solutions, and check the physical consistency of our solutions with 
the D-brane picture by looking at the distributions of their energies
and RR-charges in space. 
We also give conjectures for a few other brane configurations.
\end{abstract}

\end{titlepage}


\section{Introduction}

Following the Sen's conjectures \cite{Sentac1, Sentac2, Sentac3}, the
tachyon condensation in string theory has been attracting considerable
attention for recent a few years among many other developments in
string/M theory. In particular the tachyon condensation in open string
(field) theory provides us with an intriguing scenario \cite{Sentac3}:
all the 
string/brane physics could be reproduced via the tachyon condensation,
starting with an unstable 9-brane system, either non-BPS D9-branes in
type IIA or D9-antiD9 pairs in type IIB. Significant efforts along
this line have been made in several approaches such as the cubic
string field theory \cite{cubic}, the boundary string field theories
(BSFT) \cite{Witten1, Witten2, Shata1, Shata2, GS, Kutasov1, Kutasov2,
  Andreev, Shin, Tseytlin1, Tseytlin2} 
and the non-commutative tachyons \cite{nt1, nt2, nt3, nt4, nt5}, 
giving good amount of positive evidences for this fascinating
scenario.  

Among others, Minahan and Zwiebach proposed the field theory models
of the tachyon \cite{Zwie, MZ, MZ1, MZ2}. 
For a tachyon model of an unstable D9-brane in type IIA, they
provided the action \cite{MZ2}, 
\begin{eqnarray}
S = {\cal T}\int \!d^{10} x\; e^{-T^2/a}\left(1+(\p_\mu T)^2 +
\Half F_{\mu\nu}^2\right).
\label{MZaction}
\end{eqnarray}
This model, even with the inclusion of appropriate fermion couplings,
enjoys desirable properties as an effective theory of a non-BPS
D9-brane, such as the  
appearance of the lower dimensional D-branes as kinks or lumps with
the linear profile of the tachyon field and of discrete spectra with
an equal spacing for the fluctuations about the soliton solutions,
which are all in favor of stringy behavior of D-branes.   
This result turned out not to be an accident, for the action
(\ref{MZaction}) can actually be thought of as the two-derivative
truncation of the action obtained from the BSFT for unstable
D-branes \cite{KL, Terashima}\footnote{Precisely speaking, there 
could be an additional term, proportional to $T^2(\partial T)^2$, 
whose coefficient depends on the renormalization scheme and a certain 
choice of which can give the correct tachyon mass at the level of 
the tree level action \cite{Tseytlin1}.}.   

Up to now only the flat D-branes are constructed as kinks or lumps in
the tachyon models. So there remains the bulk of enterprises to explore
the zoology of brane configurations as solitons in the tachyon models,
such as the bound states of branes, i.e., branes within branes,
intersecting branes and branes ending on branes, etc., which is
necessary to put forward the conjecture that all the branes can be
made as kinks or lumps out of unstable $D9$-branes.  

In this paper we will make use of the action (\ref{MZaction}) and study
the solutions, such as kinks, lumps and their generalizations, which
correspond to various brane configurations in superstring theory. We
found a variety of nontrivial BPS brane configurations, including   
(i) a D8-D6 bound state, 
(ii) a (F, D8) bound state, 
(iii) a string junction,
(iv) D6-branes ending on a D8-brane, 
(v) a D6-brane suspended between two parallel D8-branes 
and (vi) their non-commutative generalizations.

The solutions obtained in this paper are expected to persist in the
inclusion of the higher derivative corrections computed in the BSFT,
as our solutions are BPS. The tensions and the RR-charges of
our solutions will be reproduced in exact agreement only when working
on the full BSFT action. Indeed the analysis of the full-fledged BSFT
action gives the exact results and will be presented in our
forthcoming paper \cite{ours}.    

\section{Brane Bound States and String Junctions} 

We shall start by constructing a few simple solutions of the action 
(\ref{MZaction})\footnote{
The BSFT action in the two-derivative truncation (again up to a term 
remarked in the previous footnote) 
is given in \cite{KL, Terashima, Kutasov2, Andreev, Tseytlin1}, 
\begin{eqnarray}
S = \sqrt{2} T_{\rm D9}\int \!d^{10} x\; 
e^{-2\pi\alpha'\widetilde{T}^2}
\left(1+4\pi\alpha'{}^2 \log 2 (\p_\mu \widetilde{T})^2 +
\pi^2 \alpha'{}^2 \widetilde{F}_{\mu\nu}^2\right).
\end{eqnarray}
If we rescale the tachyon field and the field strength in an
appropriate manner, the following identification 
\begin{eqnarray}
  a=2\alpha' \log 2,\quad
{\cal T}=\sqrt{2}T_{\rm D9}
\label{notation}
\end{eqnarray}
gives the action (\ref{MZaction}). When comparing our results 
obtained in this paper with those in the BSFT, these tuned parameters
(\ref{notation}) are to be adopted. However the two derivative
truncation employed in this paper will not provide the correct values
of energies and RR-charges expected from the well-known superstring
results. We will see the exact agreement with the expected results,
when taking into account the higher derivative corrections in
our forthcoming paper \cite{ours}. 
} 
representing the brane bound states and string junctions \cite{Dasgupta, 
Sen}. Before discussing these solutions, let us recall the basic
property of a  known solution which represents a flat D8-brane
 \cite{Kutasov2, MZ1}. 
The equations of motion for this system (\ref{MZaction}) are
\begin{eqnarray}
&&  \p_\mu ^2 T + \frac{1}{a}T
  \left(
    1-(\p_\mu T)^2 + \Half
   F_{\mu\nu}^2
  \right)=0,
\label{eomfort}\\
&&
\p_\mu 
(e^{(-T^2/a)} F_{\mu\nu})=0.
\label{eomforg}
\end{eqnarray}
The solution for the flat D8-brane is given by  
\begin{eqnarray}
  T=qx_9, \quad A_\mu=0,
\label{d8kink}
\end{eqnarray}
where a constant $q$ is fixed to be $1$. Due to the form of the
tachyon potential 
in the action (\ref{MZaction}), the kink solution is actually described by 
a linear tachyon profile, since the potential minima are located at $T=\pm
\infty$. The constant $q$ would be infinite if we
included the higher derivative corrections computed in the BSFT. In
the derivation of the BSFT action, the tachyon profile is assumed to
be a linear function, so the solution (\ref{d8kink}) is consistent
with this fact. In this paper, we will often 
assume the intuition based on the fact that the constant $q$ is
actually going to infinity when the higher derivative corrections are
included as in the BSFT. 

Now the energy of the kink solution (\ref{d8kink}) is given by
\begin{eqnarray}
  {\cal E} = 
{\cal T}
\int \!d^9x\; e^{-T^2/a}
\left(
  1+(\p_i T)^2 + \Half F_{ij}^2
\right)
=
{\cal T}\int \!d^9x\; 2e^{-T^2/a}
= 
2\sqrt{a\pi}V_{\rm D8}{\cal T},
\label{d8energy}
\end{eqnarray}
where $V_{\rm D8}$ is the volume of a D8-brane.
Therefore the energy density is localized at $x_9=0$ which represents
the position of the kink. If the constant $q$ gets larger and larger in the
inclusion of the higher derivative corrections, the localization of
the kink becomes sharper and sharper, and after all the gaussian
distribution supplied by the tachyon potential is replaced by a
delta-function .   

We can compute the Ramond-Ramond (RR) charge of the kink
solution (\ref{d8kink}) by adopting the formula 
for the Chern-Simons (CS) term given\footnote{Again 
  we have rescaled the tachyon field and the gauge fields. Thus in
  comparison with the BSFT result, we have to choose $a_1=\sqrt{2}$
  and $a_2=\sqrt{\pi/\log 2}$. } 
in \cite{KL, Terashima}:
\begin{eqnarray}
  S_{\rm CS} = \frac{\cal T}{\sqrt{2}}
 {\rm Tr} \int \! e^{-T^2/a} C \wedge \exp [a_1 F + a_2 DT],
\label{csterm}
\end{eqnarray}
where $a_1$ and $a_2$ are constant numbers.
Substituting the solution (\ref{d8kink}), one obtains the D8-brane 
charge 
\begin{eqnarray}
  S_{\rm CS} = \frac{a_2{\cal T}}{\sqrt{2}}
\int \!  C^{(9)} \wedge dT e^{-T^2/a}
=a_2\sqrt{\frac{a\pi}{2}} {\cal T}
\int_{V_{\rm D8}}\! C^{(9)},
\end{eqnarray}
which is also localized on the position of the kink, as it should be.


\subsection{D8-D6 bound state}

If we allow a constant field strength on the parent unstable D9-brane,
the constant field strength will be inherited onto the kink as well. 
This corresponds to the D8-brane containing 
lower dimensional branes on its worldvolume. These lower dimensional
branes are smeared out so that the distribution of the charge is
uniform on the D8-brane worldvolume.

Let us consider this solution more concretely.
When we turn on the constant field strength $F_{78}$, the equations of 
motion are satisfied by a linear profile for the tachyon $T$ as
\begin{eqnarray}
  T=\sqrt{1+F_{78}^2}x_9.
\label{solnaiv}
\end{eqnarray}
This solution is regarded as a D8-D6 bound state.
Actually the RR-charge is evaluated as
\begin{eqnarray}
\lefteqn{ 
 S_{\rm CS}
=\frac{a_1 a_2{\cal T}}{\sqrt{2}}\int\! d^7x\;
e^{-T^2/a}C^{(7)}_{0123456} 
\wedge (F_{78}dx_7\wedge dx_8)\wedge (\p_9T)dx_9 }\nn\\
&&\hs{30mm}+\frac{a_2{\cal T}}{\sqrt{2}}\int\! d^9x\;
e^{-T^2/a}C^{(9)}_{012345678}\wedge (\p_9T)dx_9 .
\end{eqnarray}
Therefore the constant field strength indicates that the D6-brane
charge uniformly distributed over the 78 plane, as can be seen from
the first term. The second term is the D8-brane charge which coincides
with that of the previous example. 

The energy of this bound state is evaluated as
\begin{eqnarray}
  {\cal E}_{\rm D8D6}=
\sqrt{1+F_{78}^2}\; {\cal E}_{\rm D8}.
\label{tension68}
\end{eqnarray}
This energy (\ref{tension68}) gives 
the correct ratio between the bound state energy\footnote{Even though
  we are using the two-derivative truncation of the full BSFT action,
  the result (\ref{tension68}) seems to incorporate the Born-Infeld
  result of \cite{green}. This might be due to the fact that the
  configuration we considered is supersymmetric.} of a D8- and
D6-branes and  that of a D8-brane \cite{green}. 

When we rotate the solution (\ref{solnaiv}) in the 69 plane with the
angle $\tan\theta=F_{78}$, we can obtain another expression for the
D8-D6 bound state as 
\begin{eqnarray}
  F_{78}=\mbox{const.}, \quad
T=x_9+F_{78}x_6.
\label{solbound}
\end{eqnarray}
This rotated form turns out to be useful for later use, when we study
 the non-commutative monopole in Sec.\ 4.3.

\subsection{(F, D8) bound state and string junction}

When we turn on an electric field instead of a magnetic field,
we have an (F, D8) bound state \cite{Witten, LuRoy} solution:
\begin{eqnarray}
  F_{06}=\mbox{const.},
\quad
T=\sqrt{1-F_{06}^2}\; x_9.
\label{solfd8}
\end{eqnarray}
The energy of this kink solution is calculated as 
\begin{eqnarray}
  {\cal E}_{\rm (F,D8)}
=\frac{1}{\sqrt{1-F_{06}^2}} {\cal E}_{\rm D8},
\label{fd8ene}
\end{eqnarray}
where the front factor originates from the tachyon potential
$e^{-T^2/a}$ in the action after the integration over $x_9$.  
The energy (\ref{fd8ene}) is in agreement with the ordinary
Born-Infeld analysis of the 
energy of this bound state: the Hamiltonian density of the Born-Infeld 
system of a BPS D8-brane with a constant electric field is given by 
\begin{eqnarray}
  {\cal H} = {\cal L}-\Pi^{\mu} \dot{A_\mu} = 
\frac{1}{\sqrt{1-F_{06}^2}} {\cal H}_{D8},
\end{eqnarray}
where $\Pi^\mu$ is the conjugate momentum for the gauge fields, defined
by $\Pi^\mu = \delta {\cal L}/\delta \dot{A_\mu}$ as usual.

Now we will proceed to the construction of a string junction
\cite{Dasgupta, Sen} in the tachyon model. Let us start with a string
junction on a 2-dimensional plane spanned by $x_6$ and $x_9$, and take
T-dualities along 1234578 directions. Then we arrive at the junction
composed of a D8-brane, a (F, D8) bound state and the fundamental
strings, instead of a D1-brane, a (F, D1) bound state and the
fundamental strings. Now we can utilize the D8-brane solution
(\ref{solfd8}) as a component of the junction. Hereafter we 
will call this T-dualized string junction simply as junction. 

To construct this junction solution in the tachyon model, it is useful
to refer to the construction given in \cite{Dasgupta}. 
Assuming that a constant electric field $F_{06}$ 
is turned on only in the region $x_6\geq0$ , we have the solution
(\ref{solfd8}) of the (F, D8) bound state in this region. Now in
the other region $x_6<0$, there is no electric field turned on, so we have
only a pure D8-brane instead of a (F, D8) bound state. If we connect
these two solutions at $x_6=0$, we will have a junction formed at this
point, as is similar to \cite{Dasgupta}. Due to the force balance at
the junction point, the pure D8-brane in the region $x_6<0$ must be
tilted relative to the (F, D8) bound state. Thus, in this region, we
put the ansatz 
\begin{eqnarray}
  T=q_1 x_9 + q_2 x_6  \quad (x_6<0).
 \label{soltiltd8}
\end{eqnarray}
Substituting this into the equations of motion, we obtain the
constraint 
\begin{eqnarray}
  q_1^2 + q_2^2 =1.
\end{eqnarray}
Since two kink solutions (\ref{solfd8}) and (\ref{soltiltd8}) should be
connected on a plane 
$x_6=0$, we obtain $q_1=\sqrt{1-F_{06}^2}$ and $q_2=F_{06}$. 
(Another possibility $q_2=-F_{06}$ represents the opposite orientation
of the junction.)

These two kinks are localized around the curves determined by the
equation $T=0$ : One is on $x_9=0$ $(x_6\geq0)$, and the other is on
$\sqrt{1-F_{06}^2}x_9+F_{06}x_6=0$ $(x_6<0)$. It is easy to see that 
this is the same as the configuration given in \cite{Dasgupta},
if we rotate the whole configuration on 69 plane by $\theta$ where
$\sin\theta = F_{06}$. As
claimed in \cite{Dasgupta}, we have to assume the existence of
the invisible fundamental strings stuck to the junction point. 

Note that to obtain the uniform electric field only in the half of
the 69 plane, we have to align positive electric charges on a plane 
$x_6=0$ and negative charges on $x_6=+\infty$ (assuming $F_{06}>0$). 
The electric flux covers the entire 
half of the plane, $x_6\geq 0$. However the ``effective'' electric
flux is confined only on the 
surface $\sqrt{1-F_{06}^2}x_9+F_{06}x_6=0$. This is because the
 exponential factor of the tachyon potential in front of the gauge
 kinetic term in the action (\ref{MZaction}) is
very small outside this surface. Thus the 
fundamental strings can be found only on the (F, D8) bound state. We
will make a comment on unseen fundamental strings in the discussion.

It is straightforward to generalize the above junction configuration
to the case with multiple D8-branes \cite{Kojijun}. It is simply done
by considering multiple D9-branes from the beginning. Let us consider
the junction, (0,2)-($p$,$-1$)-($-p$,$-1$), as an example. This can be 
obtained from two non-BPS D9-branes, and the solution is easily found
as 
\begin{eqnarray}
  T=
\left\{
  \begin{array}{ll}
\sqrt{1-p^2}\;x_9\unit \; &(x_6\ge 0) \\
\sqrt{1-p^2}\;x_9\unit +p\sigma_3  \; &(x_6<0) 
  \end{array}
\right. ,\quad
F_{06}=
\left\{
  \begin{array}{l}
p\sigma_3  \quad (x_6\ge 0) \\
0 \quad (x_6<0) 
  \end{array}
\right. .
\end{eqnarray}
To realize this configuration, we aligned electric charges on the 
planes $x_6=0$ and $x_6=+\infty$ appropriately for each of the 
$\sigma_3$ component of the gauge fields. 


\section{Branes Ending on Branes}

The branes ending on branes are one of the most interesting brane
configurations. For instance, it is known that D6-branes can end on a
D8-brane \cite{openp}. It is difficult to construct such
configurations in the supergravity, while neat solutions  
are available in the effective gauge theories of BPS D-branes and they 
have been known as BIons \cite{Callan, Gary}. In this section we will
look for a solution akin to BIons in the tachyon model
(\ref{MZaction}).  


\subsection{A spike (kinky-lump) solution}

An important feature of the kink solution (\ref{d8kink}) is that when
a constant $q$ becomes infinitely large in  
the inclusion of the higher derivative terms, the equation $T=0$
exactly gives the location of D-branes sharply localized on this
surface. 
Thus it is quite plausible to make 
the following ansatz for the solution representing the
branes ending on branes:
\begin{eqnarray}
T = q \left(x_9 - \frac{s}{r}\right)  ,
\label{kinky}
\end{eqnarray}
where $q$ and $s$ are constant parameters, and 
$r = \sqrt{x_6^2 + x_7^2 + x_8^2}$. 
This is  an exact analogy of the BIon which is 
given by 
$\Phi = s/r$ 
with $\Phi$ being a scalar field on the D8-brane that represents the
displacement of the D8-brane, as the equation $T=0$ with the
above ansatz (\ref{kinky}) describes the same bending $x_9=s/r$ 
of the D8-brane as that for the BIon. 
Thus the above tachyon profile \ (\ref{kinky}) gives a spike solution.
The parameter $s$ will remain a non-zero constant even in the inclusion of 
the higher derivative corrections, for it is observed that a constant
$s$ is not vanishing in the analysis of the Born-Infeld action and its
higher derivative inclusions as in \cite{Callan, Thoracius}.  

Let us substitute the above ansatz (\ref{kinky}) into the equations of
motion (\ref{eomfort}) and (\ref{eomforg}). Noticing that 
$(\p_\mu T)^2=q^2 (1+s^2/r^4)$, one can easily see that 
the constant 1 in the parenthesis of the tachyon equations of motion
(\ref{eomfort})  
is canceled if $q=1$.
This is precisely analogous to the kink solution of a D8-brane
(\ref{d8kink}). 
There is, however, another contribution $(\p_i T)^2$ ($i=6,7,8$),
coming from a nontrivial deformation of the tachyon profile. This
contribution is actually cancelled by turning on the gauge field
strength as 
\begin{eqnarray}
B_i = qs \frac{x_i}{r^3} = \p_i
\left(
  \frac{-qs}{r}
\right).
\label{solBi}
\end{eqnarray}
Here $B_i$ is a magnetic field.
This form of the magnetic field is the same as that of the BIon.
The gauge equations of motion (\ref{eomforg}) are also satisfied, 
as the linear dependence on $x_9$ in the tachyon field is irrelevant
to these equations and the contribution from the magneto-static
potential $1/r$ is easily found to be cancelled by those from the
magnetic fields on simple symmetry grounds. 

Hence we have obtained a nontrivial solution in the tachyon model
(\ref{MZaction}), which represents a D6-brane (0123459) ending on a
D8-brane (012345678), in an exact analogy with the BIon. 

The tachyon solution (\ref{kinky}) is very similar to the one
obtained in a supersymmetric sigma model with a hyper-K\"ahler target 
space  \cite{gaun}. The authors of \cite{gaun} constructed
``kinky-lump'' solution which describes branes ending on a
brane.  In their model the tachyon minima are placed in the finite 
distance instead of the infinite distance as in our case. 
Hence the kink solution is described not by a linear function as
$qx_9$ but by a hyperbolic tangent function. Also in \cite{gaun}, the
branes are ending on a two 
dimensional space, instead of a three dimensional space in our case. 
Thus they have $\log |r|$ as a deformation of the tachyon profile,
instead of $1/r$ in our solution.


\subsection{The distribution of RR-charge}

To check the physical consistency of our solution, we will look 
at the RR-charge.
Naively one might think that the distribution of the magnetic field
(\ref{solBi}) is somewhat strange, for it has no dependence
on the $x_9$ direction in which we expect the D6-branes are extended. 
If the D6-branes are ending on a D8-brane,
the D6-brane charge should be distributed on a semi-infinite plane
$x_9>0$ and $x_{678}=0$.

This can be explicitly checked by using the formula (\ref{csterm}) for
the CS term. 
Let us substitute our spike solution (\ref{kinky}) and (\ref{solBi})
into the CS term 
\begin{eqnarray}
S_{\rm CS} = \frac{{\cal T}}{\sqrt{2}}\int e^{-T^2/a}  C^{(6)}
 \wedge  \left(a_1a_2dT \wedge F + a_2^3dT \wedge dT
   \wedge   dT\right).
\label{RRform}
\end{eqnarray}
The second term $(dT)^3$ is vanishing, while the first term $dT \wedge F$
takes the form\footnote{Here we have omitted the term proportional to
  $dx_9$ which will be vanishing after the integration over
  $dx_6dx_7dx_8$, assuming that the RR 7-form is independent of 
  $x_6$,$x_7$ and $x_8$.} 
\begin{eqnarray}
dT \wedge F = q \frac{s^2}{r^4} dx_6 \wedge dx_7 \wedge dx_8.  
\end{eqnarray}
Now the D6-brane charge can be read off from
\begin{eqnarray}
\lefteqn{\frac{a_1 a_2{\cal T}}{\sqrt{2}}
\int\! d^6x \int\! dx_9\; C^{(6)}_{0123459}
\int\! dx_6 dx_7 dx_8\; e^{-T^2/a}\; q\;\frac{s^2}{r^4}} \nn\\ 
&&
=\frac{4\pi^{3/2}a_1 a_2{\cal T}}{\sqrt{2}}
\int\! d^6x \int\! dx_9\; C^{(6)}_{0123459}
 \;S(qx_9/\sqrt{a}), 
\label{d6charge}
\end{eqnarray}
where we defined a smeared step function
\begin{eqnarray}
  S(x) \equiv \Half
  \left(
    1 + \mbox{erf}(x)
  \right),
\end{eqnarray}
and erf$(x)$ is the error function defined as
\begin{eqnarray}
  \mbox{erf}(x) \equiv \frac{2}{\sqrt{\pi}}
\int_0^x \! du\; e^{-u^2}.
\end{eqnarray}
{}Thus the CS coupling (\ref{d6charge}) implies that 
the D6-brane charge is distributed only on the semi-infinite plane
$x_9>0$ ($x_{678}=0$), as we expected (here we have assumed $s>0$ for 
simplicity).    
The distribution of the D6-brane charge is sharpened when $q$ gets
larger and in the $q\to\infty$ limit the 
smeared step function $S(q x_9/\sqrt{a})$ approaches to the step
function $\theta(x_9)$.

\subsection{The Bogomol'nyi bound}

Since the form of the action (\ref{MZaction}) itself is similar to the
one of the 
 ordinary Scalar-Maxwell theory except for the overall exponential factor,
it is easy to guess the existence of the Bogomol'nyi bound 
in the tachyon model. Actually when we turn only on the tachyon and the
magnetic fields in the $x_{678}$ directions, the energy is nicely 
arranged into
\begin{eqnarray}
  {\cal E} \hs{-3mm}&&={\cal T}V_{12345}
\int \! d^4x \; e^{(-T^2/a)}
\left[
  (1-\p_9T)^2 + (\p_iT-B_i)^2
\right]\nn\\
&&\hs{10mm}+
{\cal T}V_{12345}\int \! d^4x\; 2 e^{(-T^2/a)}
\left[
  \p_9 T + \p_i (TB_i)
\right].
\label{BPSbou}
\end{eqnarray}
Thus if the configuration satisfies the BPS equations
\begin{eqnarray}
  \p_9T = 1, \quad \p_iT=B_i\quad (i=6,7,8),
\label{BPSeqn}
\end{eqnarray}
the energy
is bounded by the topological quantities:
\begin{eqnarray}
  {\cal E} \!\!\!\!\!\!\!\!&&= 
2\sqrt{\pi a}{\cal T}V_{12345}\int \!d^3x\;
\left[
  S(T/\sqrt{a})
\right]_{x_9=-\infty}^\infty
+ 2\sqrt{\pi a}{\cal T}V_{12345}\int \! d x_9\;
\int_{r=\infty}\! dS_i
\left(
  B_i S(qx_9/\sqrt{a})
\right)\nn\\
&& =
2\sqrt{\pi a}{\cal T}V_{\rm D8} 
+ 8\pi s\sqrt{\pi a}{\cal T}V_{12345}\int \! d x_9\;
S(qx_9/\sqrt{a})
.
\end{eqnarray}
The first term gives the standard kink charge
which corresponds to the energy of the D8-brane\footnote{
The first term becomes subtle at the singular point $r=0$. 
We cannot define the energy there, 
since the D8-brane surface is placed at the infinity. 
} extended along
$x_{678}$ in accord with (\ref{d8energy}). The second term is more
interesting. This provides the 
energy of the D6-brane extended along the $x_9$ axis and terminated at 
$x_9=0$, as can be seen from the smeared step function $S$ given above. 

One may notice that the BPS bound (\ref{BPSbou}) gives
precisely the same expression as that of the RR-charges
(\ref{RRform}). This 
 is similar to the case of the BPS D-branes in that 
the BPS bounds obtained from the Born-Infeld action is equal to 
the CS term for the RR-fields. This property is characteristic of the
supersymmetric D-branes, implying the plausibility of the BPS bound we
have found above. 

Before closing this section, we would like to make a comment on 
multiple BIons. As long as the BPS equations (\ref{BPSeqn}) are
satisfied, we can obtain the result similar to that we have found in
this section. Thus, for example, it is possible to get a solution
representing $N$ D6-branes stuck to a D8-brane:
\begin{eqnarray}
  T=x_9-\sum_{i=1}^N 
\frac{s^{(i)}}{|{\bf x}-\bar{\bf x}^{(i)}|}, \quad
B_i=\p_i T.
\end{eqnarray}
The parameters $s^{(i)}$ can be either positive or negative, 
depending on which the $i$-th D6-brane is elongated either to 
$x_9=+\infty$ or $-\infty$. The parameters $\bar{\bf
  x}^{(i)}=(\bar{x}_6^{(i)}, \bar{x}_7^{(i)}, \bar{x}_8^{(i)})$ 
indicates the location where the $i$-th D6-brane is stuck to the
D8-brane. The evaluation of the energy and the RR-charges gives the
expected result.


\section{A Brane Suspended Between Two Parallel Branes }

As a generalization of the solution of the previous section, we
will discuss a solution of a D6-brane suspended between two parallel
D8-branes. This brane configuration is constructed as the
'tHooft-Polyakov monopole in $SU(2)$ Yang-Mills-Higgs theory
 \cite{Diaconescu, Aki}. 

\subsection{An explicit solution}

To obtain two parallel D8-branes, we have to prepare at least two
non-BPS D9-branes. In this section we will consider only two D9-branes. 
Each BPS D8-brane is given by a kink on each D9-brane via the tachyon
condensation with the linear tachyon profile.  
We employ the following non-Abelian action with the adjoint tachyon
and the gauge fields
\begin{eqnarray}
  S= {\cal T}\; {\rm Str} \int d^{10}x e^{-T^2/a}
  \left(
    \unit + (D_\mu T)(D^\mu T)
+ \Half F_{\mu\nu} F^{\mu\nu}
  \right),
\label{actionnon}
\end{eqnarray}
where ${\rm Str}$ denotes the symmetrized trace in that we symmetrize
the $U(2)$ matrices with respect to  
$T^2$ (in $e^{-T^2/a}$), $D_\mu T$ and $F_{\mu\nu}$. We adopt the
standard normalization for the generators of the gauge group $U(2)$,
whereas the 2 $\times$ 2 unit matrix $\unit$ is not normalized in that
way.  This action (\ref{actionnon}) can be thought of as the two
derivative  truncation of the action of multiple non-BPS D9-branes
computed in the BSFT  \cite{KL, Terashima}  up to the ordering
ambiguity. 

The equations of motion derived from the above action are
\begin{eqnarray}
&&  \frac{\delta S}{\delta T}
=
\sum_{\sigma}
\left[
\left(
-2D_\mu D^\mu T 
  -\frac{2}{a}T (\unit - D_\mu T D^\mu T + \Half F_{\mu\nu}
  F^{\mu\nu})
\right)
e^{-T^2/a}
\right]=0,\\
&&  \frac{\delta S}{\delta A_\mu}
=
\sum_{\sigma}
\left[
  D_\mu
  \left(
    e^{-T^2/a}F^{\mu\nu}
  \right)+ [T, e^{-T^2/a}D_\nu T]
\right]=0.
\end{eqnarray}
Here $\sum_\sigma$ denotes the symmetrization of $U(2)$ matrices 
according to the above prescription of Str. 

Now let us look for a solution of these equations. Due to the
symmetrization, 
it is possible to find out a solution which is a precise analogue of
the one of branes ending on a brane previously discussed:
\begin{eqnarray}
  T=x_9 \unit + \Phi, \quad A_i = \epsilon_{aij} x_j \frac{W(r)}{r}
\Half \sigma_{a-5},
\end{eqnarray}
where $a,i$ and $j$ run from 6 to 8, and $\sigma_{a-5}$'s are sigma
matrices. The adjoint scalar field $\Phi$ is given by
\begin{eqnarray}
  \Phi = x_a \frac{F(r)}{r}\Half \sigma_{a-5}.
\end{eqnarray}
The scalar field $\Phi$ and the gauge fields $A_i$ in fact 
take the form of the Prasad-Sommerfield limit of the 
t'Hooft-Polyakov monopole solution,
\begin{eqnarray}
  F(r)= \frac{C}{\tanh(Cr)}-\frac{1}{r},\quad
  W(r) = \frac{1}{r} - \frac{C}{\sinh(Cr)}.
\end{eqnarray}
Note that accordingly the solution satisfies the original Bogomol'nyi
equation  $B_i+D_i \Phi=0$. 

\subsection{The energy from the  Bogomol'nyi bound}

Due to the symmetrized trace prescription, the Bogomol'nyi bound
argument goes in almost the same manner as the Abelian case in the
previous section. The energy is reorganized as
\begin{eqnarray}
  {\cal E}
&&
=
{\cal T}\;
{\rm Str} \int d^{9}x e^{-T^2/a}
  \left(
    \unit + (\p_9 T)(\p_9 T)+ (D_i T)(D_i T)
+ B_i B_i
  \right)\nn\\
&&
=
{\cal T}\;
{\rm Str} \int d^{9}x e^{-T^2/a}
  \left(
    (\unit - \p_9 T)^2 +2\p_9 T+ (D_i T+B_i)^2
-2(D_i TB_i)
  \right).
  \label{nabound}
\end{eqnarray}
Strictly speaking, this is not really the bound, for the quantities   
$(\unit - \p_9 T)^2$ and $(D_i T+B_i)^2$
are not quite the perfect square. The exponential factor of the
tachyon could possibly be appearing inside the perfect square, due to
the symmetrized trace prescription. In another word, only if the two
quantities 
\begin{eqnarray}
  \unit - \p_9 T, \quad 
D_i T + B_i
\label{quantity}
\end{eqnarray}
were commutating with the tachyon $T$, the above expression
(\ref{nabound}) would give the energy bound. Our solution actually
satisfies the BPS equations, which means nothing but the vanishing of
the above two quantities (\ref{quantity}). Thus eq.\ (\ref{nabound})
indeed provides the bound of the energy. 

Let us evaluate the energy from this expression (\ref{nabound}).
For our solution, the energy is given by the topological quantities
and we have
\begin{eqnarray}
  {\cal E} =
{\cal T}\;
{\rm Str}
\left(
 2\sqrt{\pi a}\int\! d^{8}x \;\unit
\left[S(T/\sqrt{a})\right]_{x_9=-\infty}^{\infty}\!\!\!\!
-2\sqrt{\pi a}\int \! d^5x \int\! dx_9 \int_{r=\infty}\! dS_i
\left[S(T/\sqrt{a}) B_i
\right]
  \right)\! ,
\label{susene}
\end{eqnarray}
where, for this expression to be well-defined,
we defined the error function by its Taylor expansion around the 
origin.
The first term of eq.\ (\ref{susene}) gives the energy of two parallel
D8-branes, due to the trace of the unit matrix $\unit$. 
Here the tachyon field is dominated by $T\sim x_9 \unit$, thus the 
step function of the matrix argument $T$ reduces to the ordinary 
Abelian step function.  

The evaluation of the second term of eq.\ (\ref{susene}) is non-trivial,
as it should correspond to the energy of a D6-brane 
which is not simply spread over the entire space but suspended between
two parallel D8-branes along the $x_9$ axis, as we will see below.  
To compute this term,
let us diagonalize the scalar field $\Phi$ by using the gauge symmetry
so that  
\begin{eqnarray}
  T=
\left(
  \begin{array}{cc}
x_9+C/2 & 0 \\ 0 & x_9-C/2 
  \end{array}
\right),
\label{locations}
\end{eqnarray}
where we have already expanded the function $F(r)$ around $r\sim
\infty$ and neglect the higher order terms which is dumping
exponentially 
in $r$.  This diagonal form (\ref{locations}) of the tachyon 
indicates that two parallel
D8-branes are located at $x_9=\pm C/2$.
At this gauge the magnetic field at $r\sim \infty$ also becomes
diagonal: 
\begin{eqnarray}
  B_i \sim -\Half \frac{x_i}{r^3}\sigma_3.
\end{eqnarray}
Therefore we can evaluate the energy explicitly as
\begin{eqnarray}
  {\cal E}_{\rm D6} &&=
2\sqrt{\pi a}V_{12345}{\cal T}{\rm Str}
\left[
\int\! dx_9
\left(
\begin{array}{cc}
S\left((x_9+C/2)/\sqrt{a}\right) & 0 \\ 0 & 
S\left((x_9-C/2)/\sqrt{a}\right)
\end{array}
\right)
2\pi \sigma_3
\right]\nn\\
&& = 4\pi\sqrt{\pi a}V_{12345}{\cal T}
\int \! dx_9 \;
\Bigl(S\left((x_9+C/2)/\sqrt{a}\right)-
S\left((x_9-C/2)/\sqrt{a}\right)\Bigr).
\end{eqnarray}
The step function properly accounts that the energy density of a
D6-brane  is localized only on the line segment $-C/2 < x_9 < C/2$. 
Thus this is consistent with the proposed D-brane configuration.

The RR-charge of this solution can be evaluated by using the formula
(\ref{csterm}) for the CS term. One ends up with an appropriate 
distribution of the RR-charges. In particular we can manifestly see 
the D6-brane charge localized on the line segment on the $x_9$ axis. As
discussed in  Sec.\ 3.3, the CS term in this case again turns out to
be the same as the BPS bound energy, though up to the ordering
ambiguity. 

\subsection{Non-commutative monopole}

The non-commutativity on the worldvolume of D-branes are equivalent
to turning on the background NS-NS $b$-field  \cite{hull}. This
constant 
$b$-field can be thought of as a constant magnetic field on the
D-brane, so this leads us to the brane bound states in Sec.\ 2. 
Therefore, in terms of the brane configuration, 
the non-commutative monopole can be interpreted as a D6-brane ending
on a 
D8-D6 bound state  \cite{AkiKoji}. Since we have already studied
the D6-brane ending on a D8-brane as well as a D8-D6 bound state, as
in (\ref{kinky}) and (\ref{solBi}) and in (\ref{solnaiv})
respectively, it is very easy to combine these solutions together. 
For the $U(1)$ non-commutative monopole \cite{nm2, nm2, nm3}, the
corresponding solution is  
\begin{eqnarray}
  T=x_9-\frac{s}{r}+b_{78}x_6, \quad B_i =
  \frac{sx_i}{r^3}+b_{78}\delta_{i6},
\end{eqnarray}
where $b$ is the background b-field. All the analysis of the energy
and the Bogomol'nyi bound are going well, reproducing the expected
results of the energy of the non-commutative monopole.  
Also the generalization to the $U(2)$ non-commutative monopole
\cite{AkiKoji, nm4, nm5, nm6} is trivial.


\section{Conclusions, Discussions and Future Directions}

Let us summarize our results. We obtained various brane configurations
as classical solutions of the tachyon model
(\ref{MZaction}) of Minahan and Zwiebach, which include  
(i) a D8-D6 bound state,
(ii) a (F, D8) bound state, 
(iii) a string junction,
(iv) D6-branes ending on a D8-brane, 
(v)  a D6-brane suspended between two parallel D8-branes
 and (vi) their non-commutative generalizations.
We computed the energies and the RR-charges of our solutions, and
found that they showed appropriate distributions in space 
which was expected from the corresponding D-brane pictures.
Also we found that the solutions (iv), (v) and (vi) satisfy the BPS
equations, saturating the energy bounds. There is a trivial
generalization of our results (i) -- (vi) to the lower dimensional
case, though by starting with the lower dimensional analogue of the
tachyon model (\ref{MZaction}).  

We have not checked whether the energies and the RR-charges of our
solutions precisely match the ones in string theories, though we have
found their distributions 
 to be in nice agreements with the expected results. However it is
anticipated that the energies and the RR-charges will have the correct 
values only when we
include the higher-derivative corrections to the action
(\ref{MZaction}). We expect that the qualitative forms of our
solutions will not be modified by the inclusion of the higher
derivative corrections, as our solutions are supposed to preserve some
fractions of supersymmetries in
the spacetime\footnote{It might be interesting to investigate how
  we can see supersymmetries explicitly in the tachyonic field theory,
  as analyzed in \cite{Suyama}.}. In our forthcoming paper
\cite{ours}, we will study  
the higher derivative effects by making use of the BSFT action 
explicitly. We will find the perfect agreement of the energies and
the RR-charges with the expected string theory results.
Furthermore, we are going to find more variety of solutions such as 
a D$p$-D$(p\!-\!4)$ bound state  \cite{Douglas}, (F, D($p\!-\!2$))
bound states ending 
on D$p$-brane and a dielectric branes  \cite{Emparan,
  Myers}.  

There are many other interesting brane configurations, other than
those considered in this paper. We will make brief comments on some of
them.  

\begin{itemize}
\item
Nahm construction of a D8-brane from multiple D6-branes
 \cite{Diaconescu}. 

To obtain a D6-brane in the tachyon model, we have to prepare at least
two D9-branes. The D6-brane 
solution is given by the ABS construction  \cite{Kutasov2} as 
$T=q\sigma_{i-5} x_i$ ($i=6,7,8$). As in \cite{Dasgupta}, 
we know that the D8-brane
on which D6-branes are ending can be realized as a solution of the
Nahm equation (which was shown to be a BPS equation on the effective
theory of D6-branes)  
\begin{eqnarray}
  \p_9 \Phi_i = \epsilon_{ijk}\Phi_j \Phi_k,
\label{nahm}
\end{eqnarray}
where $\Phi_i$ are the adjoint scalars on the D6-branes. A
solution for the $SU(2)$ case is given by
$\Phi_i=-\sigma_i/x_9$. We need at least two 
D6-branes to have a non-trivial solution of the Nahm equation
(\ref{nahm}), as is obvious. 
It in turn means that we must have at least four unstable D9-branes. 
It is conceivable to think of the adjoint scalar $\Phi_i$ as the
deformation of the tachyon profile, as we have done in our construction
of branes ending on branes. Now we have a conjecture for the tachyon
profile as 
\begin{eqnarray}
  T
= q(\sigma_i x_i \otimes \unit - \sigma_i \otimes \Phi_i)
= q(\sigma_i x_i \otimes \unit + \sigma_i \otimes \sigma_i/x_9),
\end{eqnarray}
where the gauge field will not be necessary.

\item
Two D6-branes intersecting with each other, sharing 4+1 dimensional 
worldvolume. 

This configuration requires at least two D9-branes. Our conjecture for
the solution is 
\begin{eqnarray}
  T=q
  \left[
    \sigma_1
    \left(
      x_6-s\frac{x_4}{x_4^2+x_5^2}
    \right)
+
    \sigma_2
    \left(
      x_7+s\frac{x_5}{x_4^2+x_5^2}
    \right)
+\sigma_3 x_8
  \right].
\end{eqnarray}
Two D6-branes are oriented along the 45 and 67 planes respectively,  
while sharing the  spacetime 
along 01239. Our conjecture is based on the following argument: Let us
first prepare a single D6-brane as 
$T=q\sigma_{i-5}x_i$. Then the fluctuation about this tachyon profile
can be taken into account as $T=q\sigma_{i-5}(x_i-\Phi_i)$. The other
D6-brane will be realized by turning on the adjoint scalars with 
\begin{eqnarray}
  \Phi_6 + i\Phi_7 = \frac{s}{x_4+ix_5}.
\end{eqnarray}
This is consistent with the supersymmetric (=holomorphic) membrane
embedding in the spacetime, 
\begin{eqnarray}
  zw=s
\end{eqnarray}
where $z\equiv x_6+ix_7$ and $w\equiv x_4+ix_5$.

\end{itemize}

The final remark is concerning the fundamental strings in the closed
string vacuum, which was discussed in references
\cite{Yi,Hori,Senf}. In Sec.\ 2.2, we studied the string junction in
which the fundamental string should
have been appearing as a component of the three-pronged string, though
we could not find the confined flux in our solution. 
In addition we have to note that it seems impossible to construct 
the fundamental strings ending on a D8-brane, originally studied in
\cite{Callan}, in our tachyon model, though one might think we could
do so similarly to the construction of D6-branes ending on a D8-brane
in Sec.\ 3. This might be related to the fact that the fundamental
strings cannot be seen easily in our model. We will be faced with the
same difficulty also in the case of the (F, D6) bound state ending on
a D8-brane or suspended between two parallel D8-branes. These are
simply dyonic BIons  \cite{Gary}, but it seems hard to construct these
configurations in our tachyon model.  

We leave these issues for the future work.



\vs{10mm}
\noindent
{\large \bf Acknowledgments}

K.\ H.\ would like to thank J.\ Gauntlett for valuable
discussions. S.\ H.\ is grateful to N.\ Sasakura for discussions. 
K.\ H.\ and S.\ H.\ were supported in part by the Japan Society for
the Promotion of Science. This research was supported in part by
the National Science Foundation under Grant No.\ PHY99-07949.

\newcommand{\J}[4]{{\sl #1} {\bf #2} (#3) #4}
\newcommand{\andJ}[3]{{\bf #1} (#2) #3}
\newcommand{\AP}{Ann.\ Phys.\ (N.Y.)}
\newcommand{\MPL}{Mod.\ Phys.\ Lett.}
\newcommand{\NP}{Nucl.\ Phys.}
\newcommand{\PL}{Phys.\ Lett.}
\newcommand{\PR}{ Phys.\ Rev.}
\newcommand{\PRL}{Phys.\ Rev.\ Lett.}
\newcommand{\PTP}{Prog.\ Theor.\ Phys.}
\newcommand{\hep}[1]{{\tt hep-th/{#1}}}

\end{document}